\documentclass[12pt]{article}
\usepackage{latexsym}
\usepackage{epsfig}

\usepackage{graphicx}
\usepackage{epstopdf}

\usepackage{epsfig,amssymb,amsmath,amscd}

\newcommand{\bq}{\begin{equation}}
\newcommand{\eq}{\end{equation}}
\newcommand{\bqa}{\begin{eqnarray}}
\newcommand{\eqa}{\end{eqnarray}}
\newcommand{\ben}{\begin{enumerate}}
\newcommand{\een}{\end{enumerate}}
\newcommand{\bc}{\begin{center}}
\newcommand{\ec}{\end{center}}
\newcommand{\bqb}{\begin{eqnarray*}}
\newcommand{\eqb}{\end{eqnarray*}}

\def\pr#1#2#3{Phys. Rev. ${\bf{#1}}$, #2 (#3)}

\def\pl#1#2#3{Phys. Lett. ${\bf{#1}}$, #2 (#3)}

\def\np#1#2#3{Nucl. Phys. ${\bf{#1}}$, #2 (#3)}

\def\jmp#1#2#3{J. Mod. Phys. ${\bf{#1}}$, #2 (#3)}

\begin{document}
\pagenumbering{arabic}
\thispagestyle{empty}
\def\thefootnote{\fnsymbol{footnote}}
\setcounter{footnote}{1}

\vspace*{2cm}
\begin{flushright}
July 3, 2018\\
 \end{flushright}
\vspace*{1cm}

\begin{center}
{\Large {\bf W polarization in $e^+e^-$, gluon-gluon and $\gamma\gamma \to W t\bar b $ 
for testing the top quark mass structure and the presence of final interactions.}}\\
 \vspace{1cm}
{\large F.M. Renard}\\
\vspace{0.2cm}
Laboratoire Univers et Particules de Montpellier,
UMR 5299\\
Universit\'{e} de Montpellier, Place Eug\`{e}ne Bataillon CC072\\
 F-34095 Montpellier Cedex 5, France.\\
\end{center}

\vspace*{1.cm}
\begin{center}
{\bf Abstract}
\end{center}

We analyze the $Wt\bar b$ production processes and especially the rate of
longitudinal $W$ polarization sensitive to a possible top quark mass 
scale dependence and to the presence of strong final state $Wt$
interactions for example generated by a dark matter environment.
We give illustrations for the three processes
$e^+e^-$, gluon-gluon and $\gamma\gamma \to W t\bar b $. 

\vspace{0.5cm}

\def\thefootnote{\arabic{footnote}}
\setcounter{footnote}{0}
\clearpage

\section{INTRODUCTION}

In previous papers we have made an analysis of the $Z$ longitudinal
polarization in the three processes 
$e^+e^-$, gluon-gluon and $\gamma\gamma \to Z t\bar t $, \cite{eettZ, ggttZ}. 
We have checked that, after the well-known gauge cancellation, the rate of
longitudinal $Z$ polarization, equivalent to the rate of $G^0 t\bar t $
production \cite{equiv}, is directly sensitive to any modification of the top quark
mass. We have then shown that a scale dependent top quark mass
mentioned in \cite{trcomp,CSMrev}, for
example generated by substructures, \cite{comp, Hcomp2,Hcomp3,Hcomp4,partialcomp}, 
may immediately lead to
observable effects on this rate. On another hand, new interactions
between heavy particles, for example generated by a dark matter (DM)
environment \cite{DMmass, DMexch}, may lead to final state interactions between $Z$ and
top quarks and then also to strong modifications of the $Z_L$ rate.\\

We now want to see if similar modifications of the SM predictions could
be generated in the charged sector.
We will also first check, in the
three considered processes, the equivalence at high energies of $W^- t\bar b $ production with the $G^-t\bar b$ one in the pure SM case.
We will then illustrate the modifications of the $W^-_L$ rates which
appear when we replace $m_t$ by a scale dependent $m_t(s)$. 
The presence of the small bottom quark mass is negligible.
Using the same test functions as in the $Ztt$ cases we will finally discuss 
the effects of final state interactions between $W$ and top quark.\\
The comparison with the previous $Ztt$ cases allows us to conclude that
the $W^- t\bar b $ processes may be almost as interesting as the 
$Ztt$ ones and that their analyzis may be complementary and contribute to the determination of the basic structure of the underlying dynamics responsible for
such possible SM modifications.\\

Contents: Section 2 contains three subsctions respectively devoted to 
the $e^+e^-$, gluon-gluon and $\gamma\gamma \to Z t\bar t $
processes, presents the basic SM diagrams and illustrates the locations of the
considered modifications corresponding to the chosen explicit expressions
for the new dynamics.
Conclusions are summarized in Section 3.

\section{ANALYSES OF $Wt\bar b$ PRODUCTION PROCESSES}

\subsection{$e^+e^- \to W t\bar b $}

The Born SM diagrams are given in Fig.1. It is well-known that the
$E_W/m_W$ behaviour of the components of the $W_L$ amplitude 
cancel when adding all diagrams, leaving, up to $m^2_W/E^2_W$ corrections,
a contribution proportional to $m_t$ (and a negligible $m_b$) 
equivalent to the Goldstone production $G^-t\bar b$ amplitude with
the left and right couplings

\bq
 c^L={em_t\over\sqrt{2}s_Wm_W}~~~~~~~~c^R=-~{em_b\over\sqrt{2}s_Wm_W}~.
\eq
\noindent
This equivalence is shown in Fig. 2a,b for $\sqrt s=5$ TeV and
$\theta_W={\pi\over6}$ and ${\pi\over2}$.
The computation of $G^-t\bar b$ production is done with similar
diagrams as in Fig.1 replacing  the external $W^-$ line by
a $G^-$ one.\\
This is the starting point of our study of the $m_t$ scale dependence.\\
If it occurs like in the neutral $Ztt$ case, an $m_t(s)$ behaviour

\bq
m_t(s)=m_t{(m^2_{th}+m^2_0)\over (s+m^2_0)}
\eq
\noindent
would immediately reflect in the $W_L$ rate

\bq
R_L={\sigma(W_L t\bar b)\over \sigma(W_T t\bar b)+\sigma(W_L t\bar b)}
\eq
as shown in Fig.3a,b with $m_0=2,4$ TeV, curves $m2$ and $m4$..\\
The result is somewhat weaker than in the $e^+e^-\to Zt\bar t$ case,
essentially at low angles because of the presence of $W_T$ emission 
from the $e^{\pm}$ lines which have no Goldstone counterpart.\\
As in the previous cases we have also looked at a possible final
state interaction between heavy particles (here only $Wt$), for 
example generated by a DM environment, modifying the $W_L t\bar b$
amplitudes by the $[1+C(s_{Wt})]$ "test factor"
with
\bq
C(x)=1+{m^2_{t}\over m^2_0}~ln{-x\over (m_W+m_t)^2} ~~, \label{Fs}
\eq
\noindent
for the subenergy $x=s_{Wt}$  and $m_0=0.5$ TeV, like in \cite{DMexch}
and in the $Zt\bar t$ case.\\
We can appreciate in Fig.4a,b the similarity
of the effects (although relatively smaller) as compared to those of the 
$e^+e^-\to Zt\bar t$ case (with only the $Wt$ final interaction, curve $DMW$,
or by adding the $Gt\to Wt$, curve $DMWG$).\\ 
Experimental possibilities for such processes can be found in \cite{ee}.
\\

\subsection{$gg \to W t\bar b $}

We do the same analysis for the gluon-gluon process whose SM diagrams
are given in Fig.5.\\
The difference with the $e^+e^-$ case is essentially the absence of 
direct W emission from the initial state.\\
The corresponding results are shown in Fig.6a,b; 7a,b and 8a,b.\\
Apart from a somehat different angular dependence the qualitative
effects of an $m_t$ scale dependence or of final state interactions
are rather similar to the ones in $Ztt$.\\
The LHC possibilities can be found in \cite{Contino, Richard}.\\

\subsection{$\gamma\gamma \to W t\bar b $}

This process may be particularly interesting because of the additional different
type of diagrams with self gauge boson couplings as shown in Fig.9.\\
The results are shown in Fig.10a,b; 11a,b and 12a,b.\\
The shapes of the distributions are also slightly different from
the gluon-gluon ones and could provide independent tests of the considered new physics effects.\\
The possibilities with photon-photon collisions are reviewed in
\cite{gammagamma}.\\

\section{CONCLUSION}

In this paper we have extended our previous study of the possible effects of a 
scale dependent top quark mass $m_t(s)$ and of the presence
of final state interactions between heavy particles. These effects may originate
from the top quark subsctructure and/or a special interaction with a dark matter
environment. For this type of search we had previously considered $Z$ polarization in $Zt\bar t$ production processes.
We have now enlarged our study to $W t\bar b $ processes in order to allow a comparison of the charged and of the neutral sectors.\\
The rates of $Z_L$ and of $W_L$ production are indeed directly sensitive
to these new dynamical features. In the $W t\bar b $ case the absence of $m_b$
effects and the restriction of the new final interactions to the $(Wt$) couple leads to relatively smaller
but nevertheless still important observable effects with typical shapes.
We have made illustrations by using 
arbitrary test functions not originating from a well-defined model. Their purpose
is just to show that the measurement of the
longitudinal polarization rates may be very instructive for determining
the nature of the underlying dynamics responsible for these effects.\\
To summarized them, a decrease of $m_t(s)$ leads immediately to a
corresponding decrease of the $Z_L$ and $W_L$ rates. On another hand a final ($Zt$) or ($Wt$) interaction (for example generated by dark matter environment
specific to heavy particles) can also directly lead to a modification of these rates.\\
Dedicated experimental studies should then be done in order to transform our present illustrations into modifications of observable quantities.\\

\newpage

\begin{figure}[p]
\vspace{-0cm}
\[
\hspace{-2cm}\epsfig{file=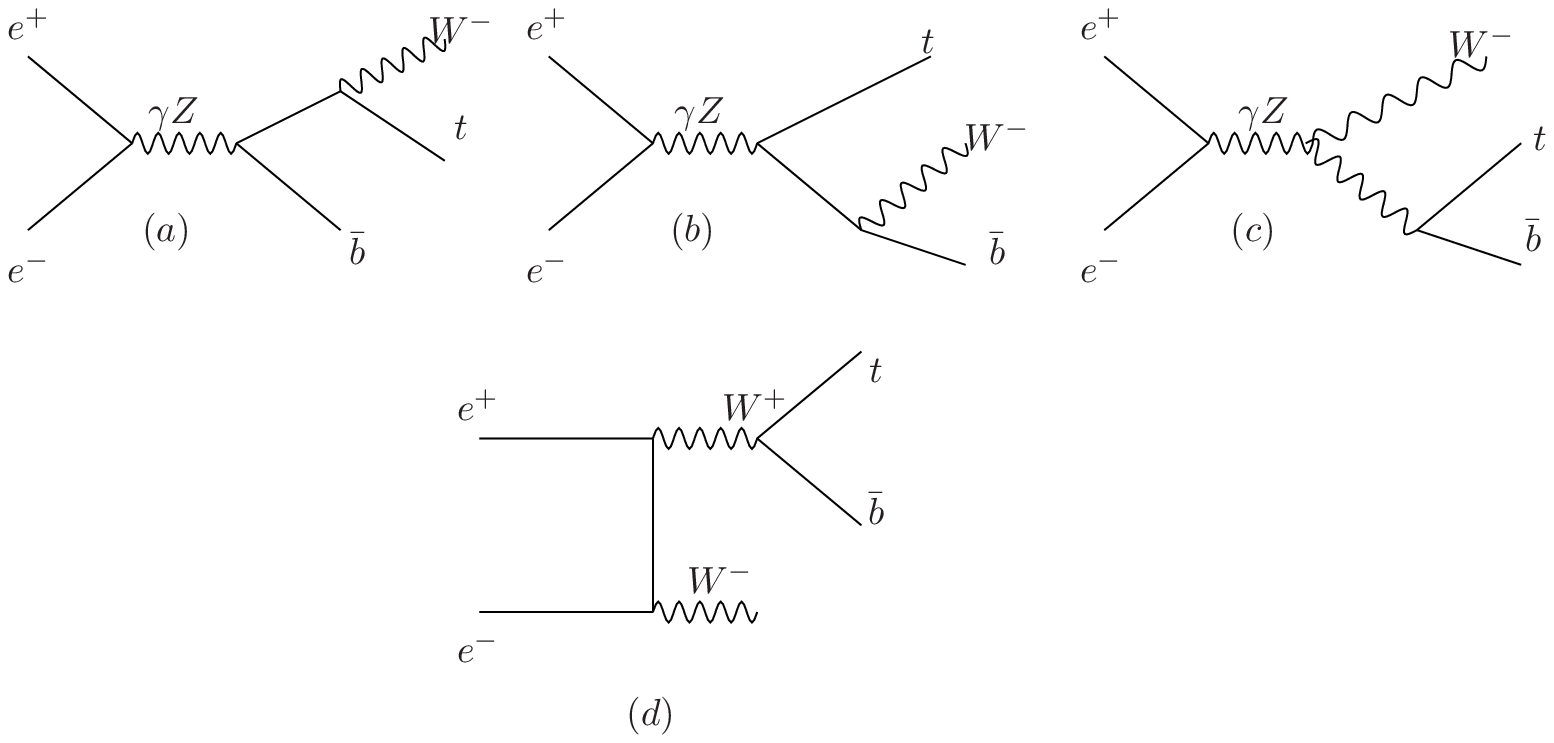 , height=9.cm}
\]\\
\vspace{1cm}
\caption[1] {SM diagrams for $e^+e^-\to W^-t\bar b$ Born amplitudes.}
\end{figure}
\clearpage

\begin{figure}[p]
\vspace{-0cm}
\[
\hspace{-2cm}\epsfig{file=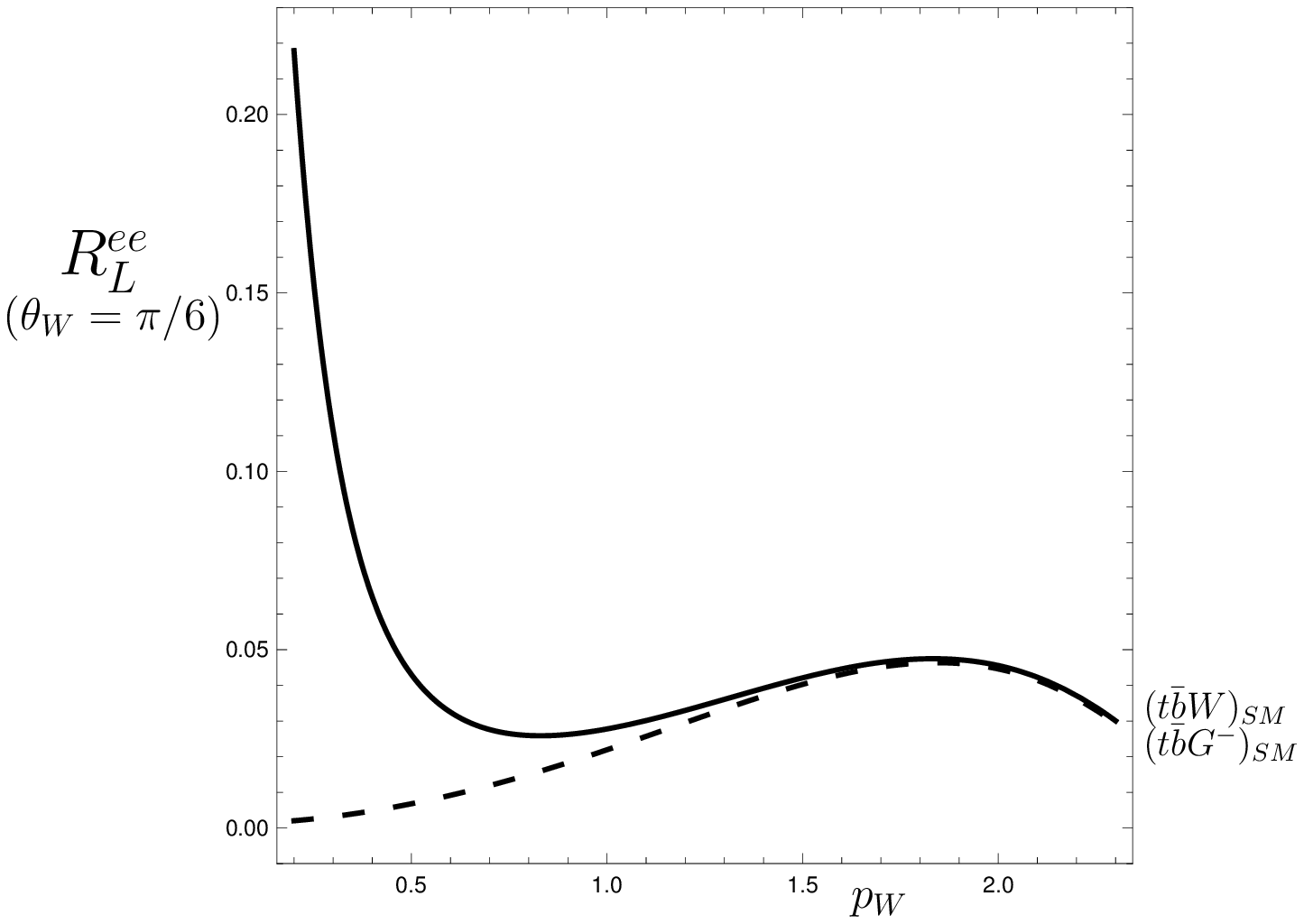 , height=9.cm}
\]\\
\vspace{0.5cm}
\[
\hspace{-2cm}\epsfig{file=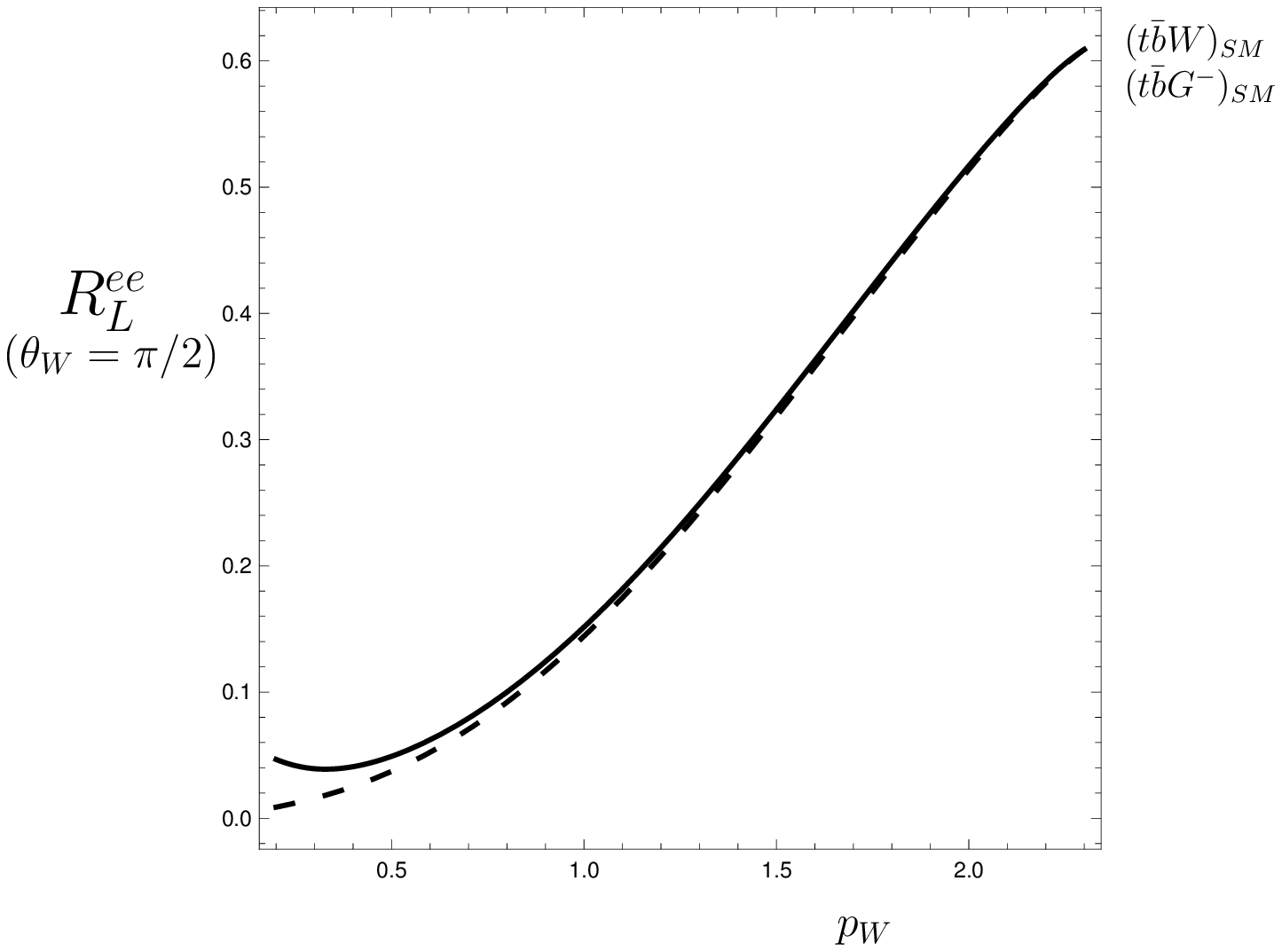 , height=9.cm}
\]\\
\vspace{-1cm}
\caption[1] {$e^+e^- \to W_L t\bar b $ ratio compared to the Goldstone case.}
\end{figure}
\clearpage

\begin{figure}[p]
\vspace{-0cm}
\[
\hspace{-2cm}\epsfig{file=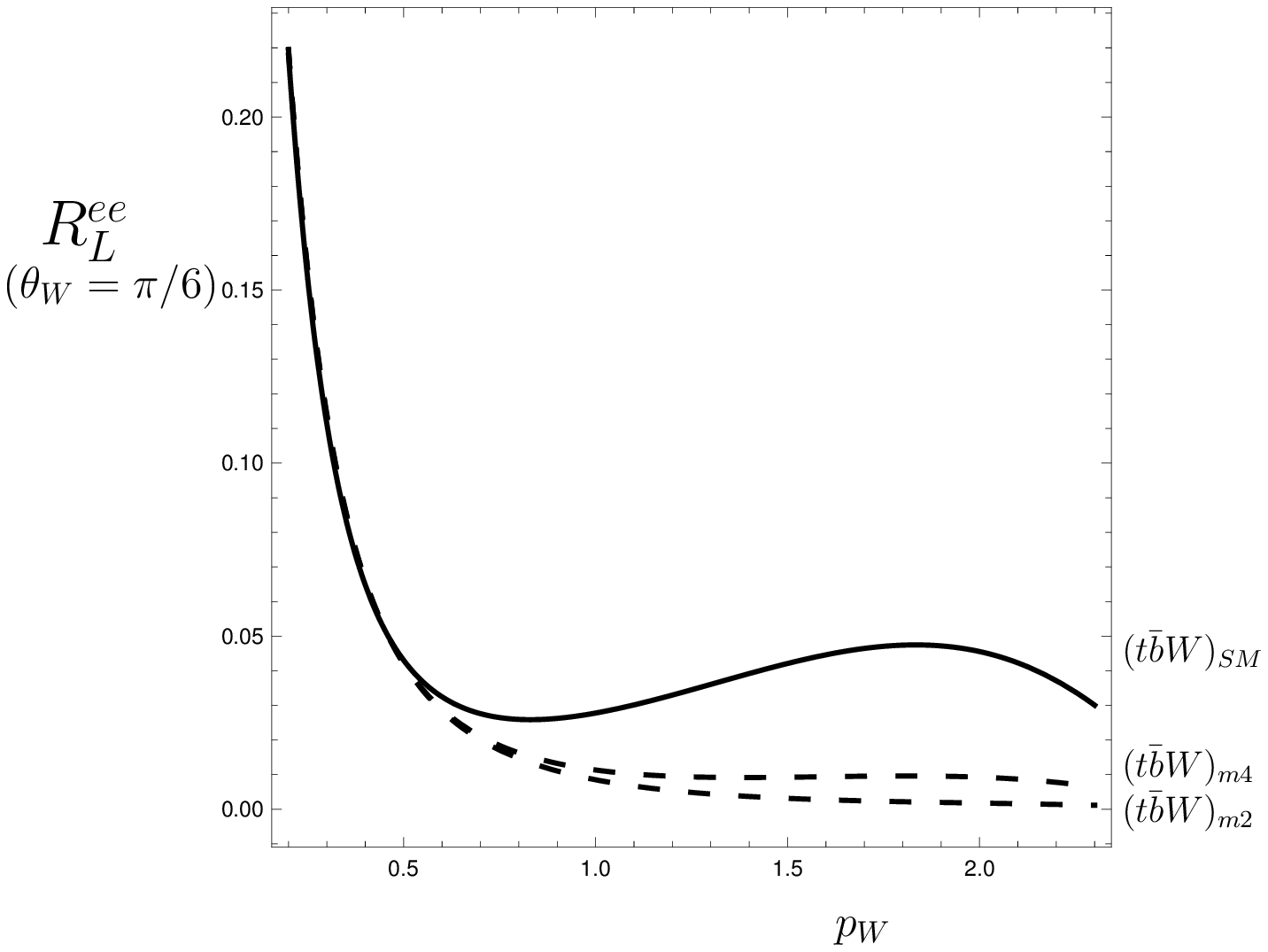 , height=9.cm}
\]\\
\vspace{0.5cm}
\[
\hspace{-2cm}\epsfig{file=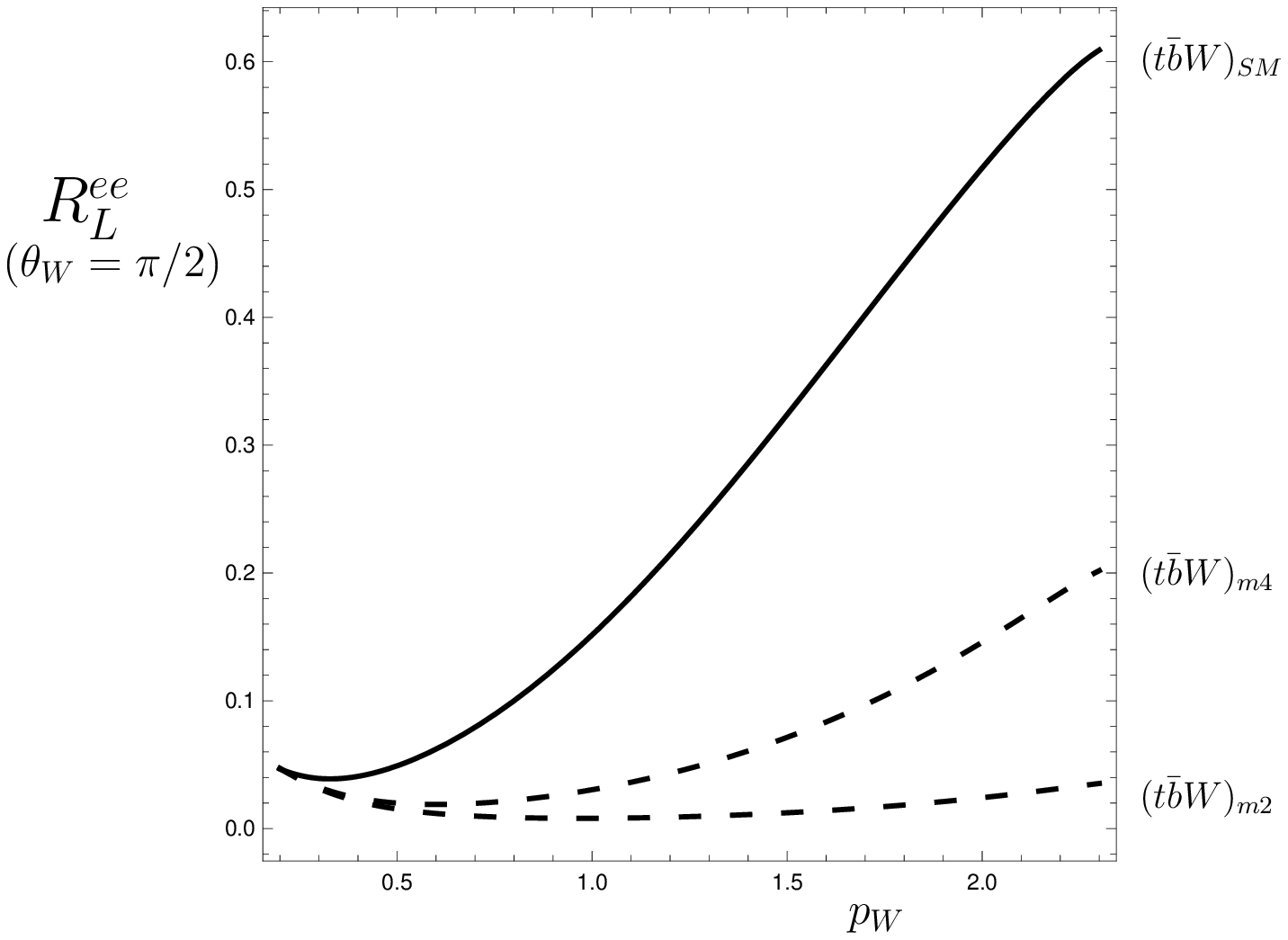 , height=9.cm}
\]\\
\vspace{-1cm}
\caption[1] {$e^+e^- \to W_L t\bar b $ ratio for 2 cases of scale dependent top mass compared to the SM case.}
\end{figure}
\clearpage

\begin{figure}[p]
\vspace{-0cm}
\[
\hspace{-2cm}\epsfig{file=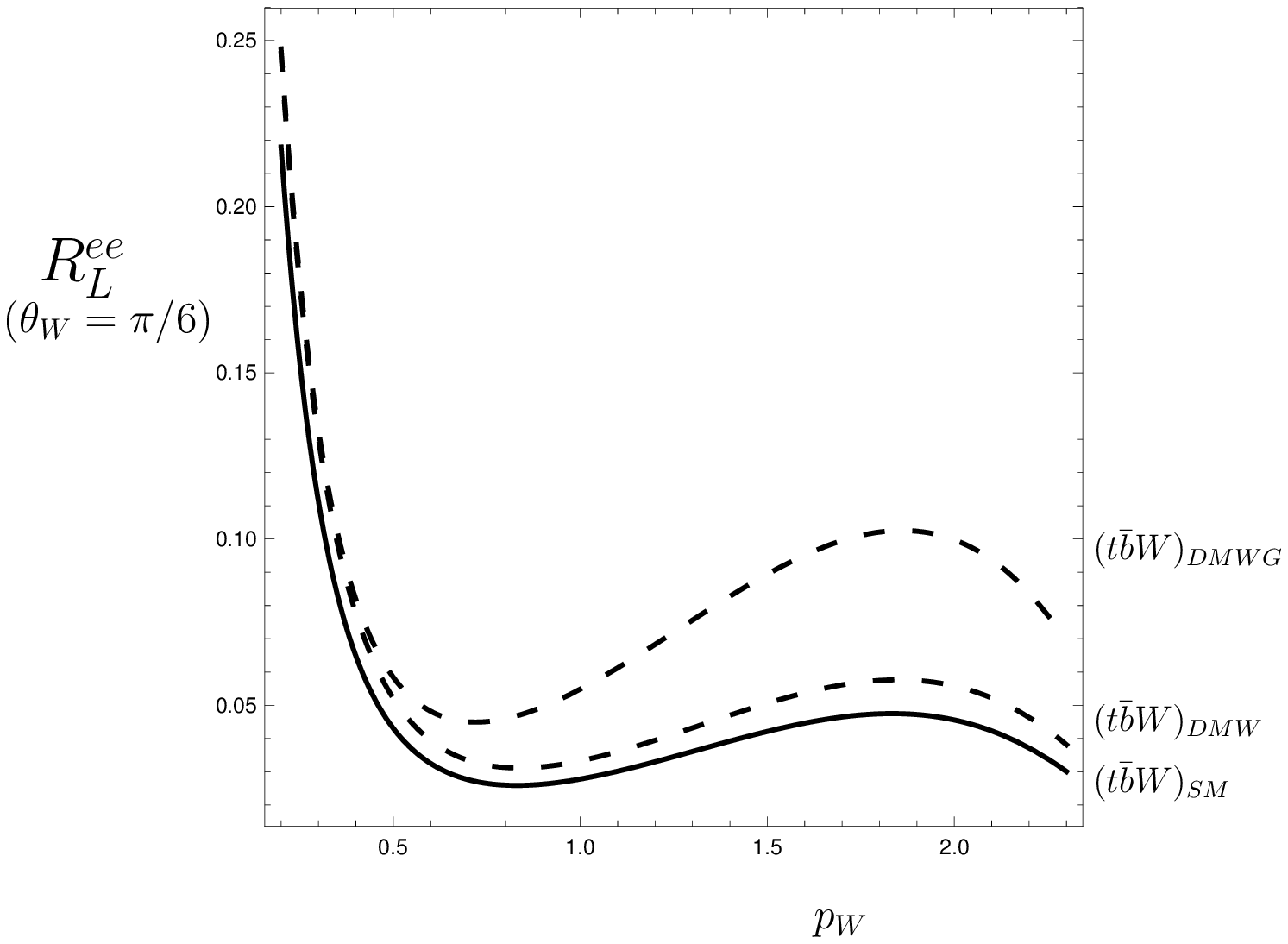 , height=9.cm}
\]\\
\vspace{1cm}
\[
\hspace{-2cm}\epsfig{file=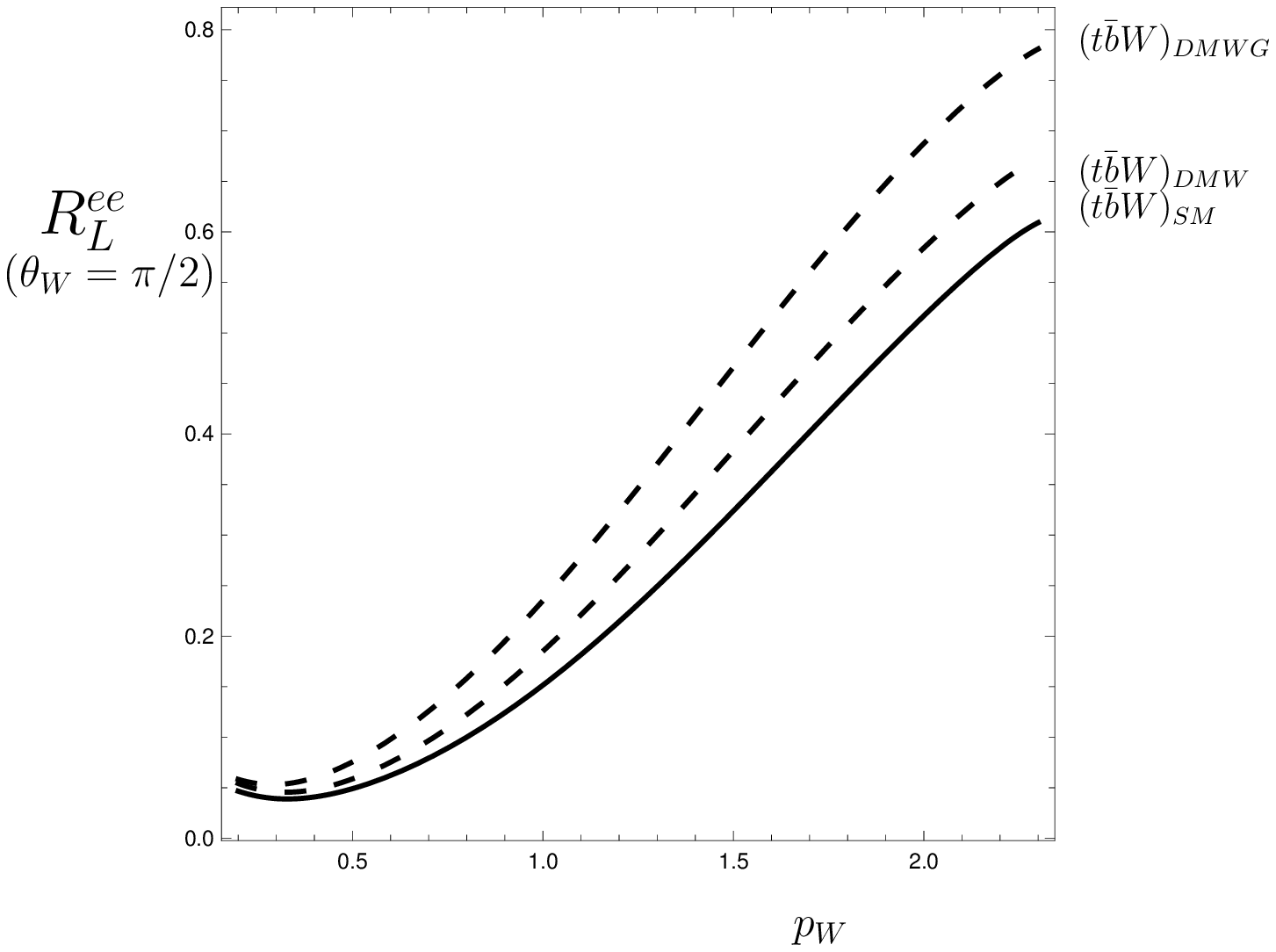 , height=9.cm}
\]\\
\vspace{-1cm}
\caption[1] {$e^+e^- \to W_L t\bar b $ ratio for 2 cases of Dark Matter final state interactions.}
\end{figure}
\clearpage

\begin{figure}[p]
\vspace{-0cm}
\[
\hspace{-2cm}\epsfig{file=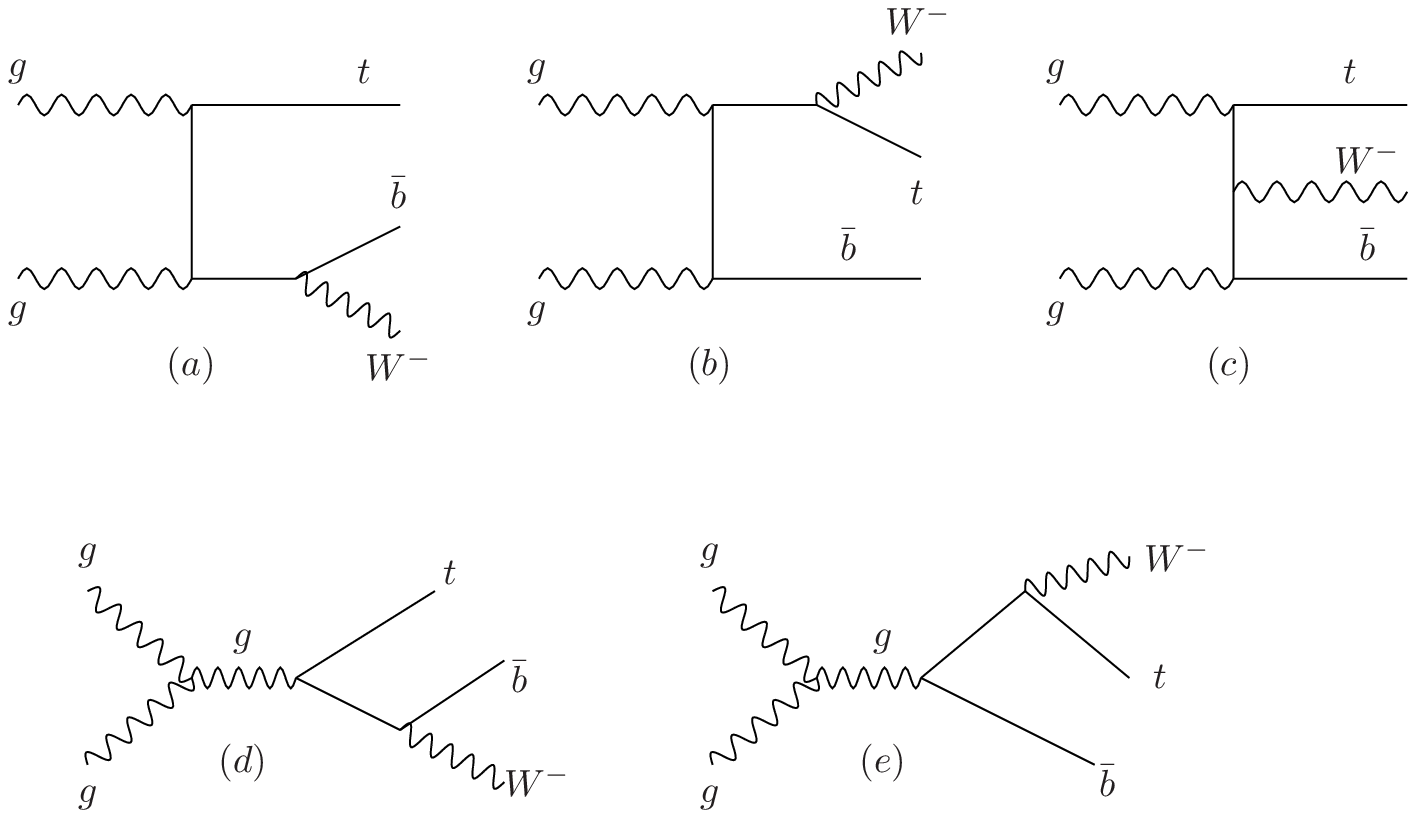 , height=9.cm}
\]\\
\vspace{2cm}
\caption[1] {SM diagrams for $gg\to W^-t\bar b$ Born amplitudes.}
\end{figure}
\clearpage

\begin{figure}[p]
\vspace{0cm}
\[
\hspace{-2cm}\epsfig{file=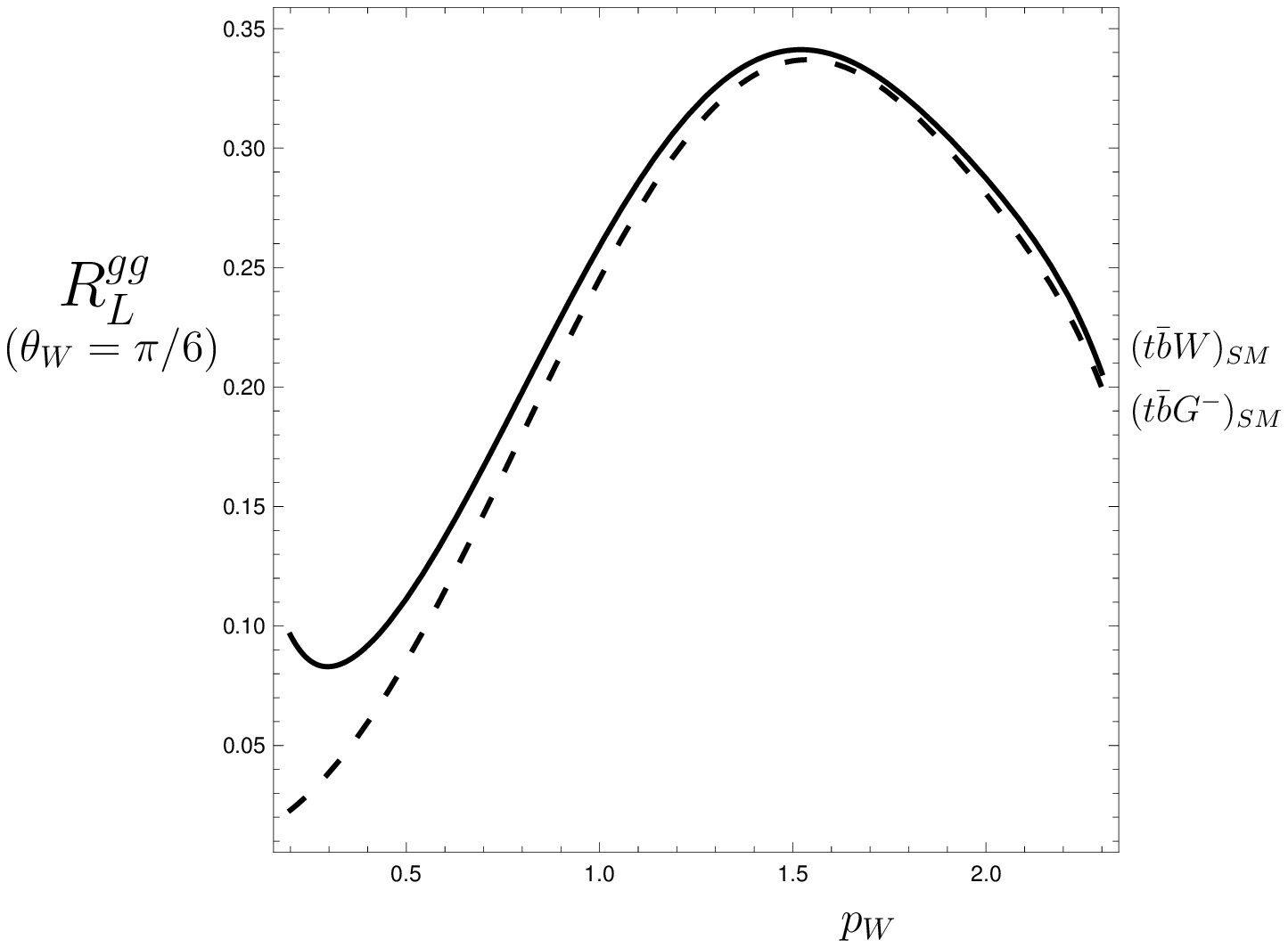 , height=9.cm}
\]\\
\vspace{0cm}
\[
\hspace{-2cm}\epsfig{file=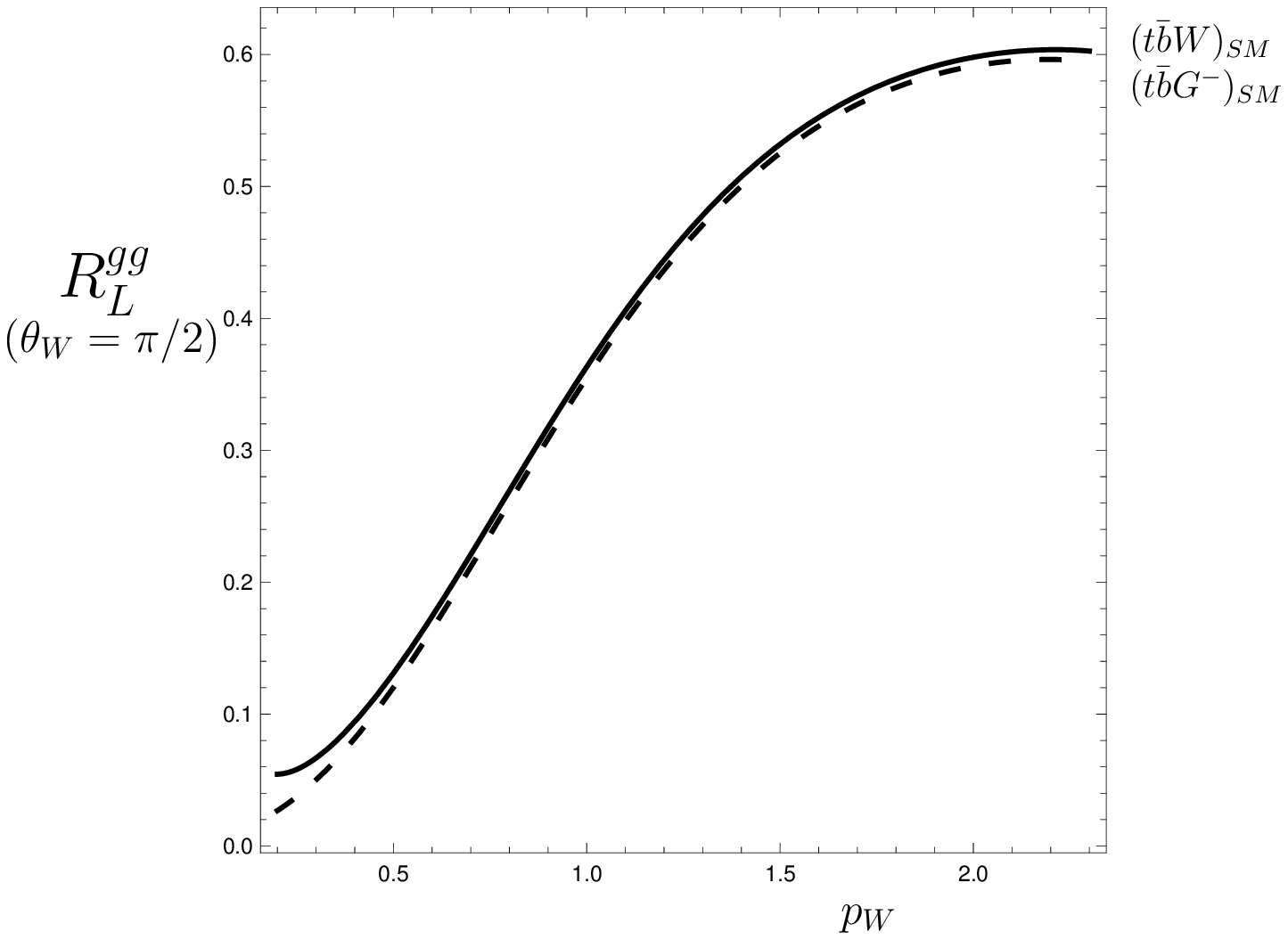 , height=9.cm}
\]\\
\vspace{-1cm}
\caption[1] {SM $gg\to W_L t\bar b$ ratio compared to the Goldstone case.}
\end{figure}

\clearpage

\begin{figure}[p]
\vspace{-1cm}
\[
\hspace{-2cm}\epsfig{file=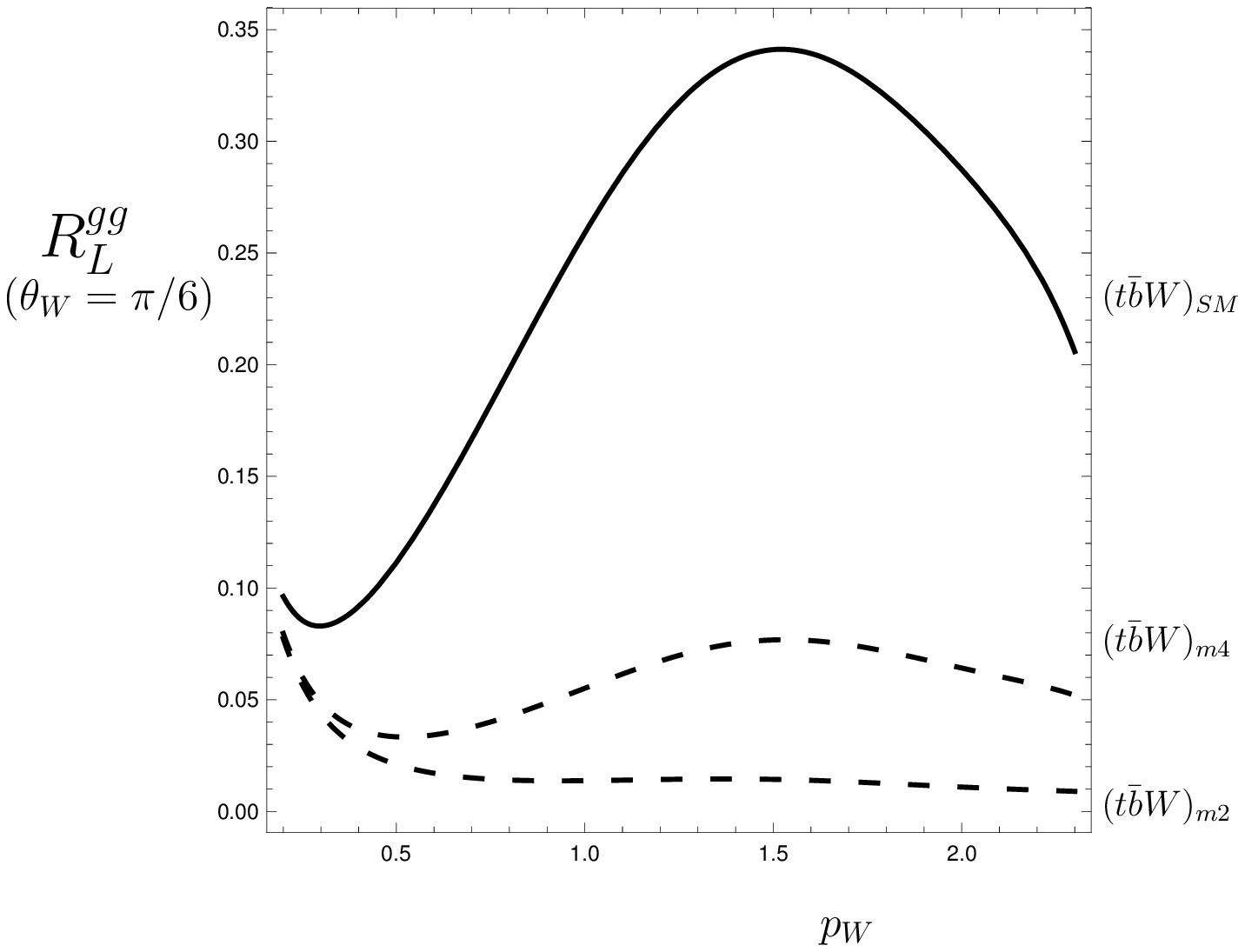 , height=9.cm}
\]\\
\vspace{0.5cm}
\[
\hspace{-2cm}\epsfig{file=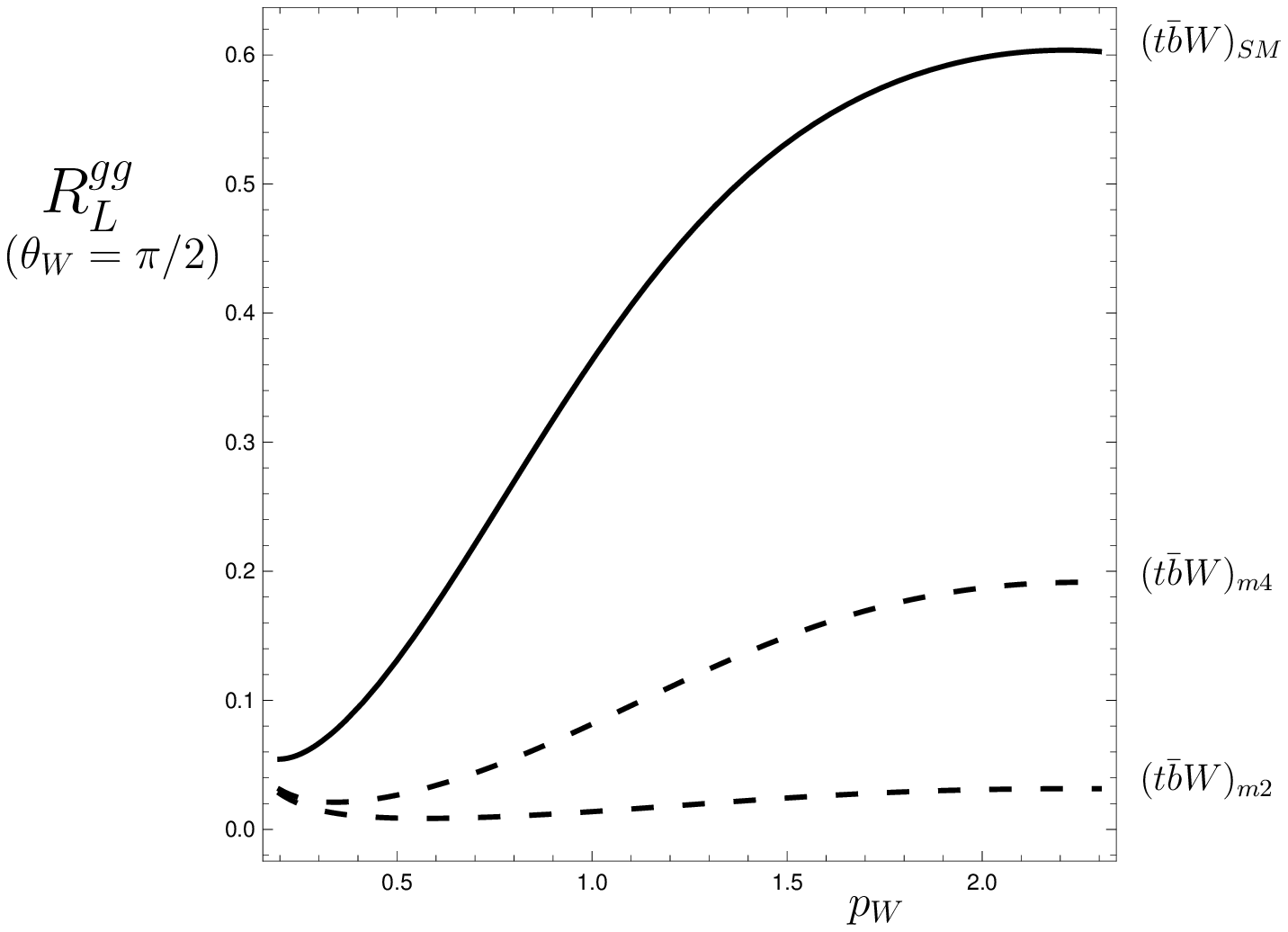 , height=9.cm}
\]\\
\vspace{-1cm}
\caption[1] {$gg \to W_L t\bar b$ ratio for 2 cases of scale dependent top mass compared to the SM case.}
\end{figure}
\clearpage

\begin{figure}[p]
\vspace{-1cm}
\[
\hspace{-2cm}\epsfig{file=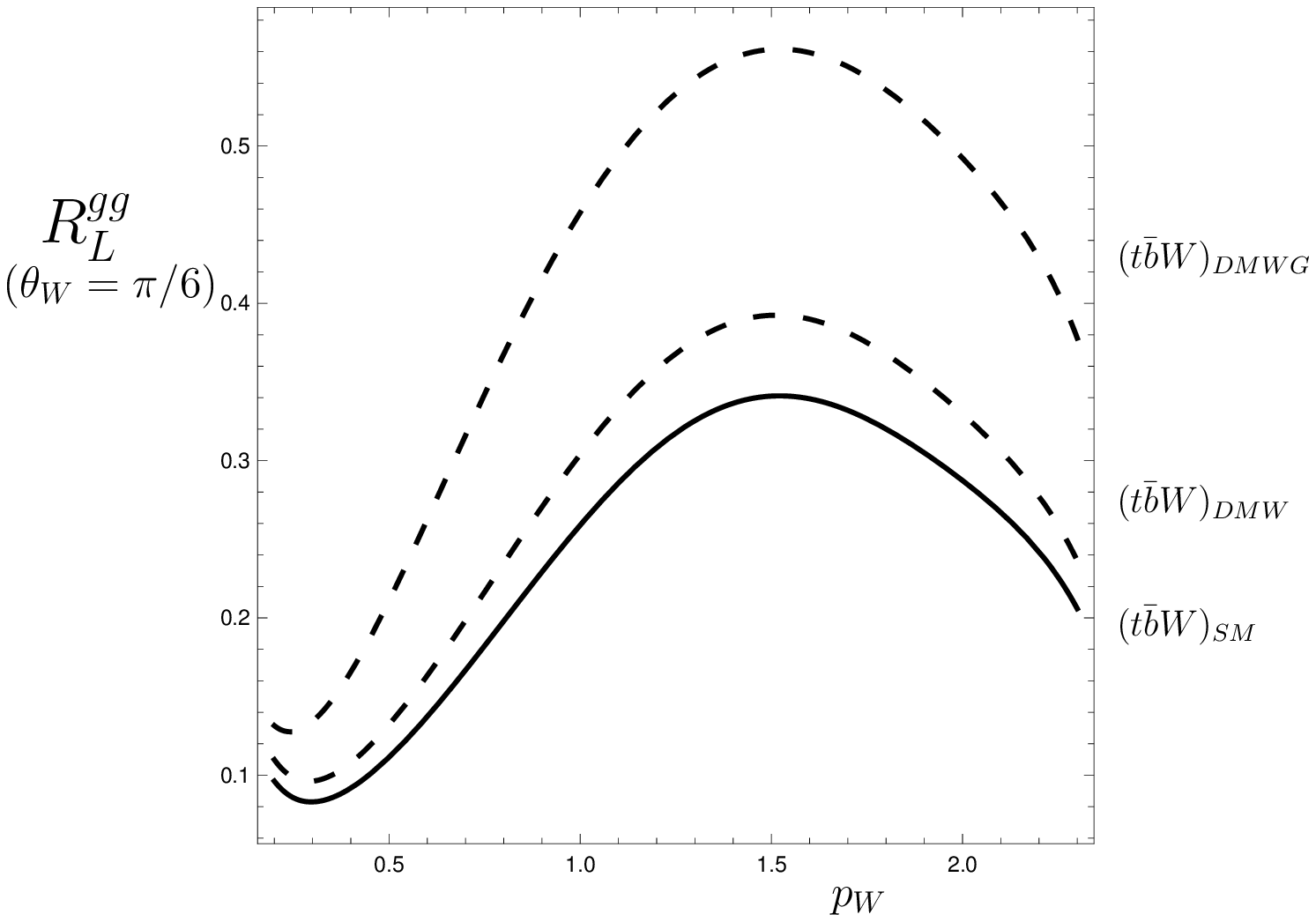 , height=9.cm}
\]\\
\vspace{0.5cm}
\[
\hspace{-2cm}\epsfig{file=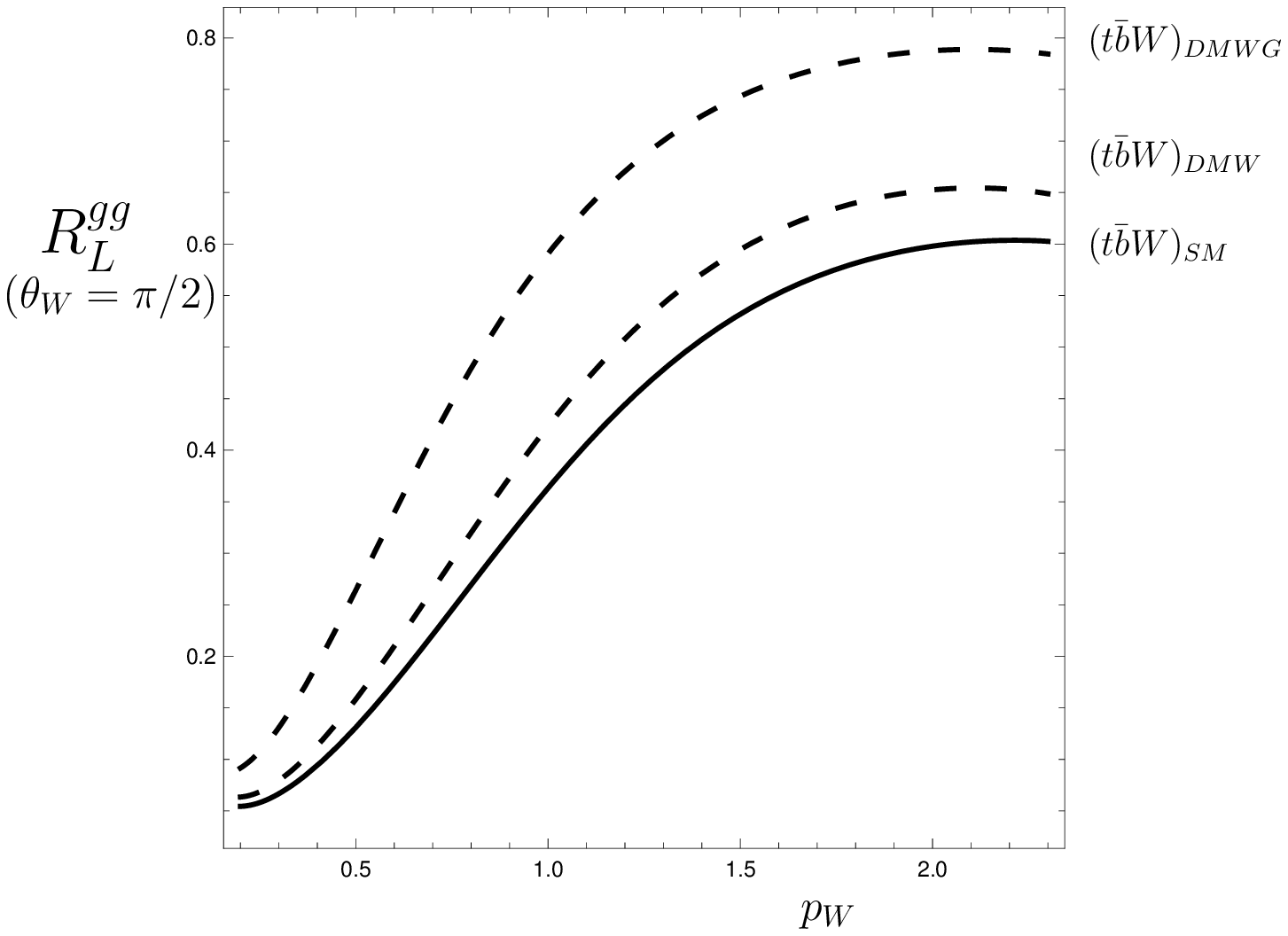 , height=9.cm}
\]\\
\vspace{-1cm}
\caption[1] {$gg \to W_L t\bar b$ ratio for 2 cases of Dark Matter final state interactions.}
\end{figure}

\clearpage

\begin{figure}[p]
\vspace{-1cm}
\[
\hspace{-2cm}\epsfig{file=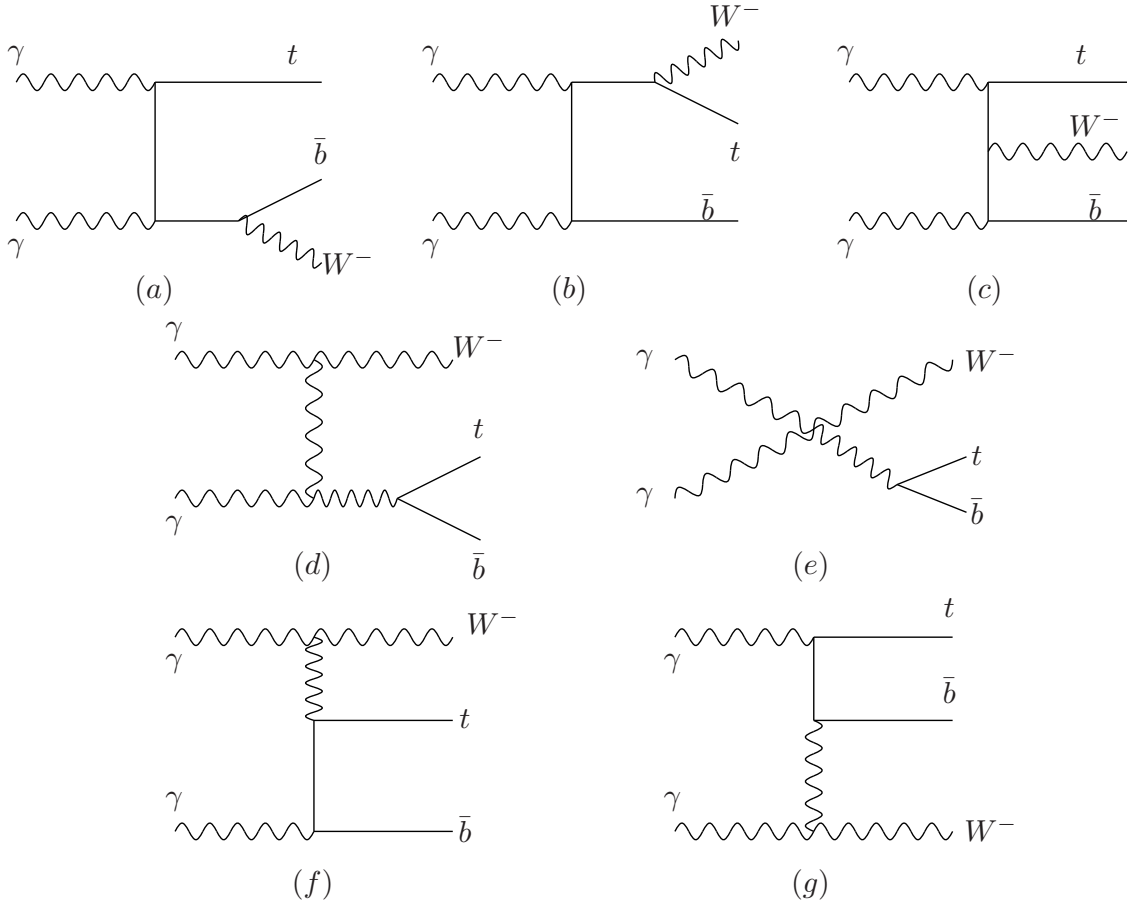 , height=12.cm}
\]\\
\vspace{1cm}
\caption[1] {SM diagrams for $\gamma\gamma\to W^-t\bar b$ Born amplitudes.
Internal weavy lines represent both virtual gauge and goldstone bosons.}
\end{figure}
\clearpage

\begin{figure}[p]
\vspace{0cm}
\[
\hspace{-2cm}\epsfig{file=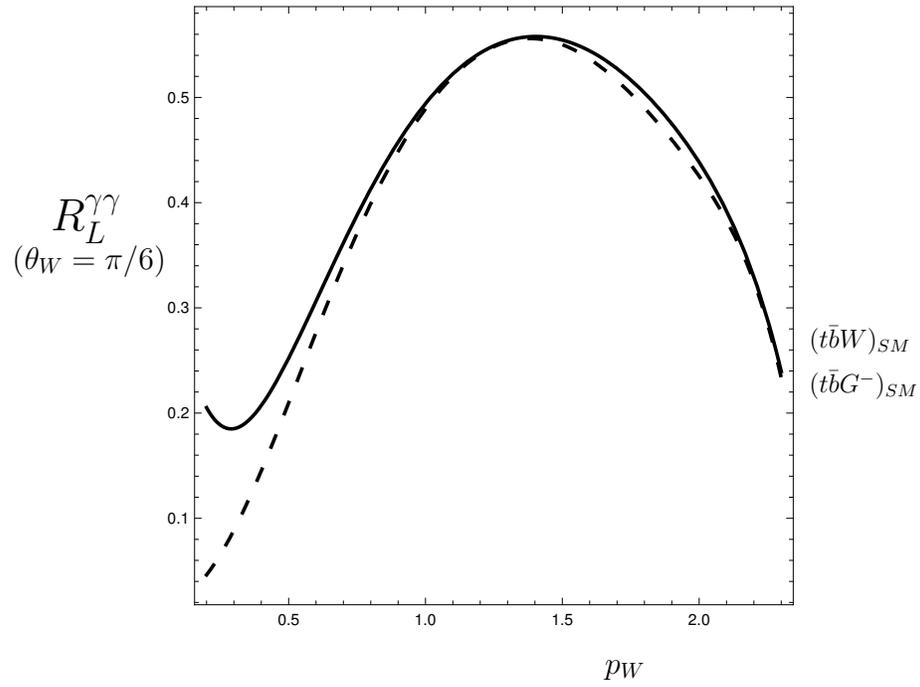 , height=9.cm}
\]\\
\vspace{0cm}
\[
\hspace{-2cm}\epsfig{file=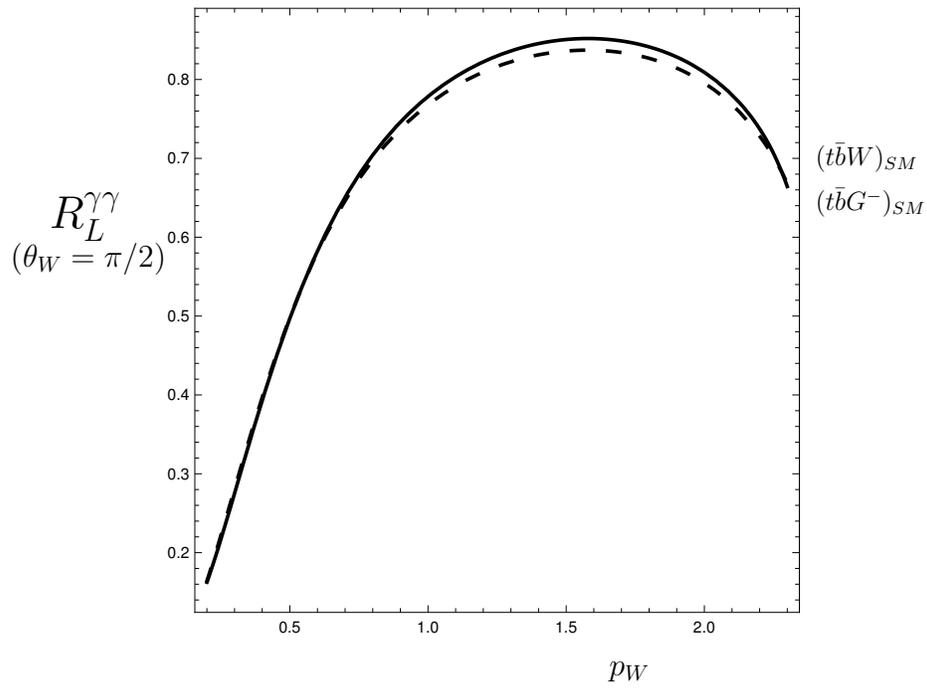 , height=9.cm}
\]\\
\vspace{-1cm}
\caption[1] {SM $\gamma\gamma\to W_L t\bar b$ ratio compared to the Goldstone case.}
\end{figure}

\clearpage

\begin{figure}[p]
\vspace{-1cm}
\[
\hspace{-2cm}\epsfig{file=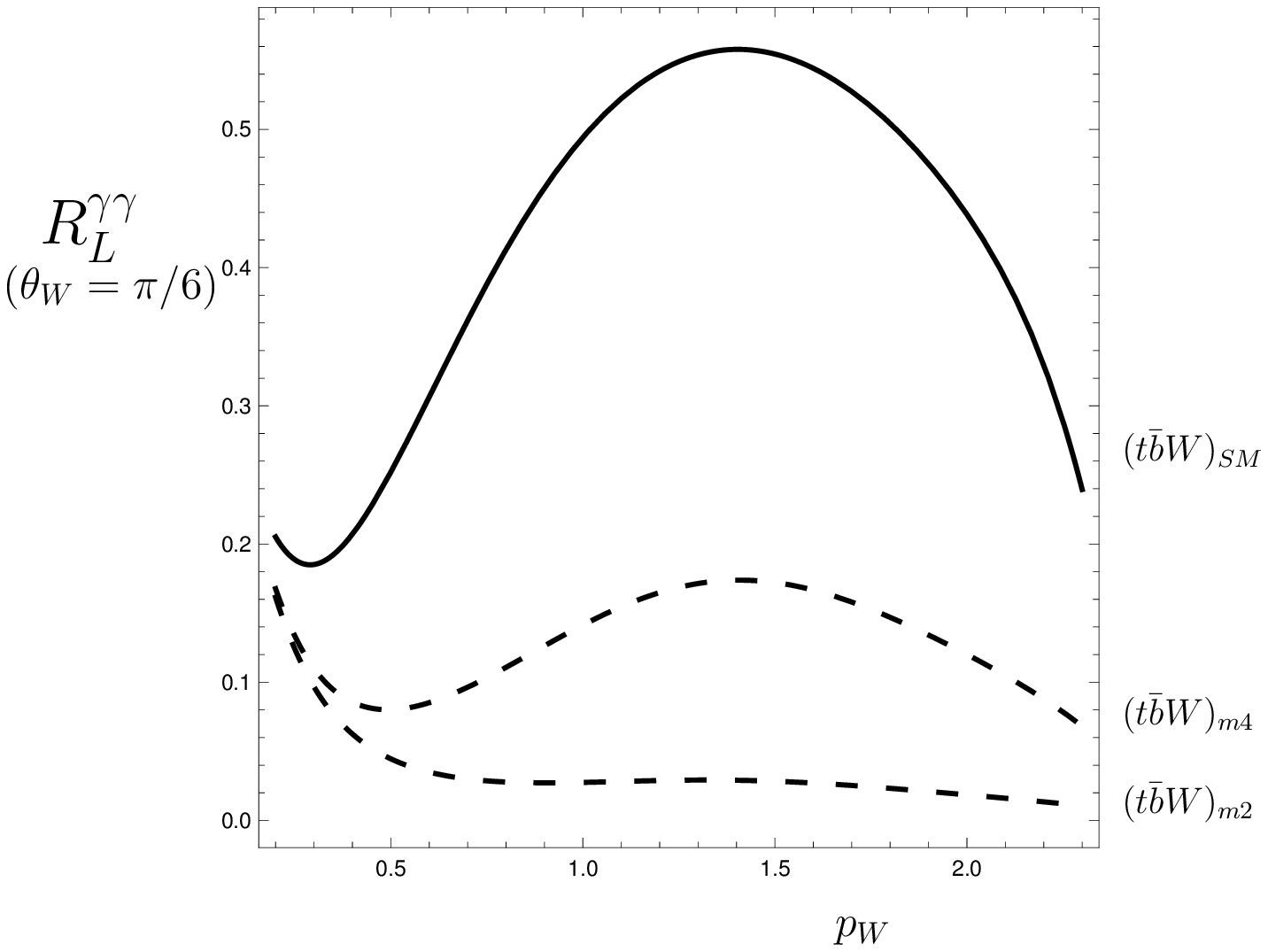 , height=9.cm}
\]\\
\vspace{0.5cm}
\[
\hspace{-2cm}\epsfig{file=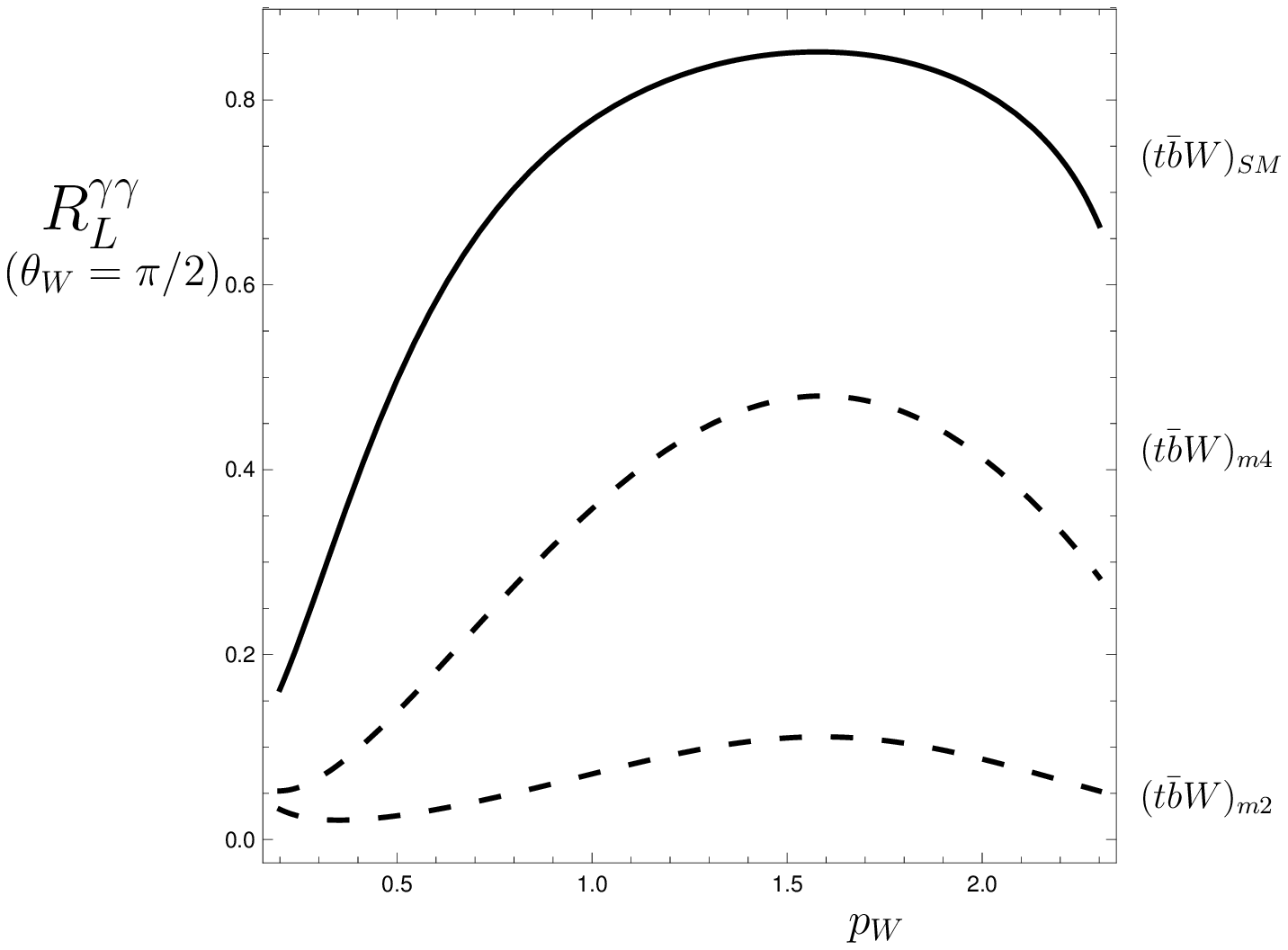 , height=9.cm}
\]\\
\vspace{-1cm}
\caption[1] {$\gamma\gamma\to W_L t\bar b$ ratio for 2 cases of scale dependent top mass compared to the SM case.}
\end{figure}
\clearpage

\begin{figure}[p]
\vspace{-1cm}
\[
\hspace{-2cm}\epsfig{file=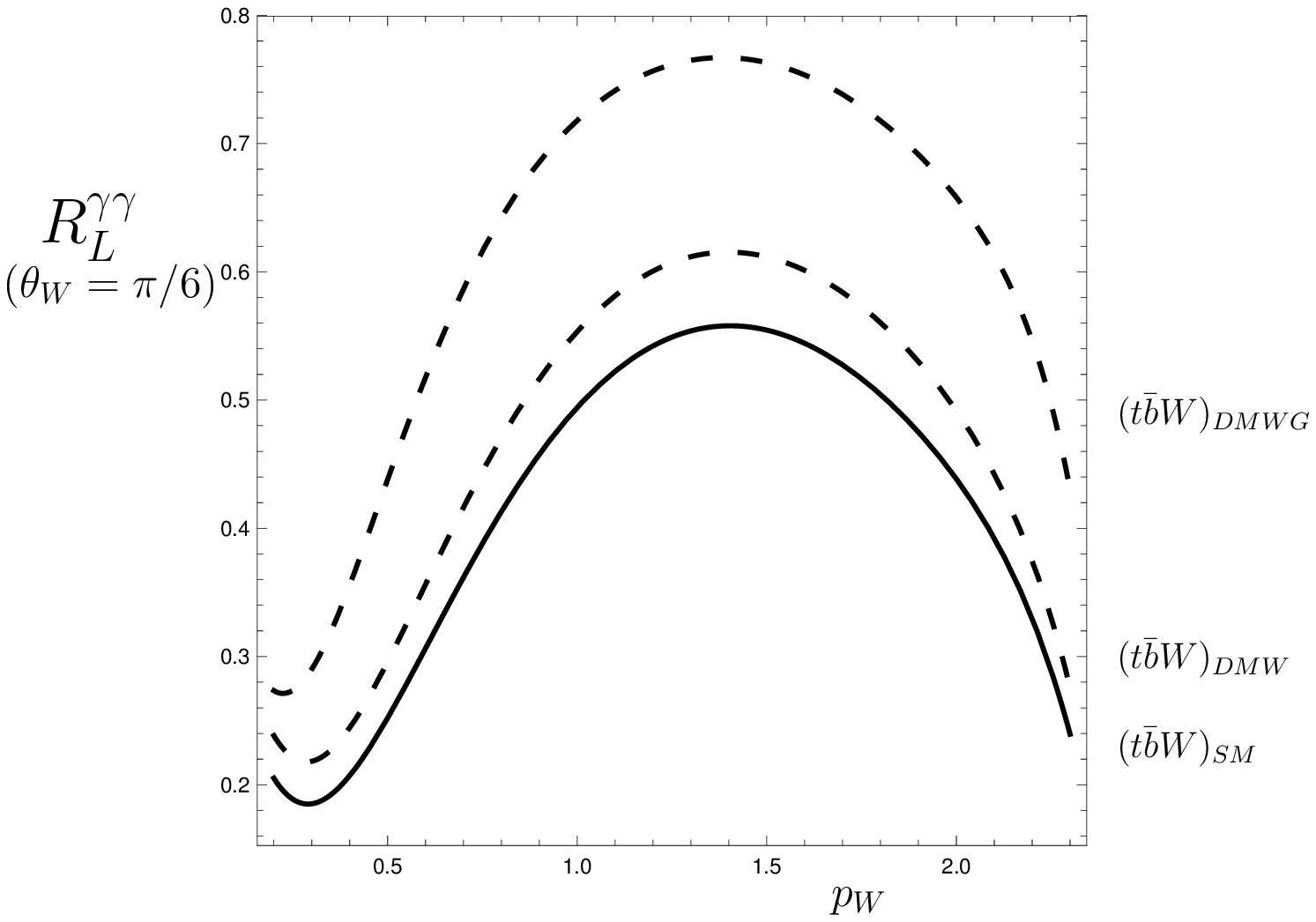 , height=9.cm}
\]\\
\vspace{0.5cm}
\[
\hspace{-2cm}\epsfig{file=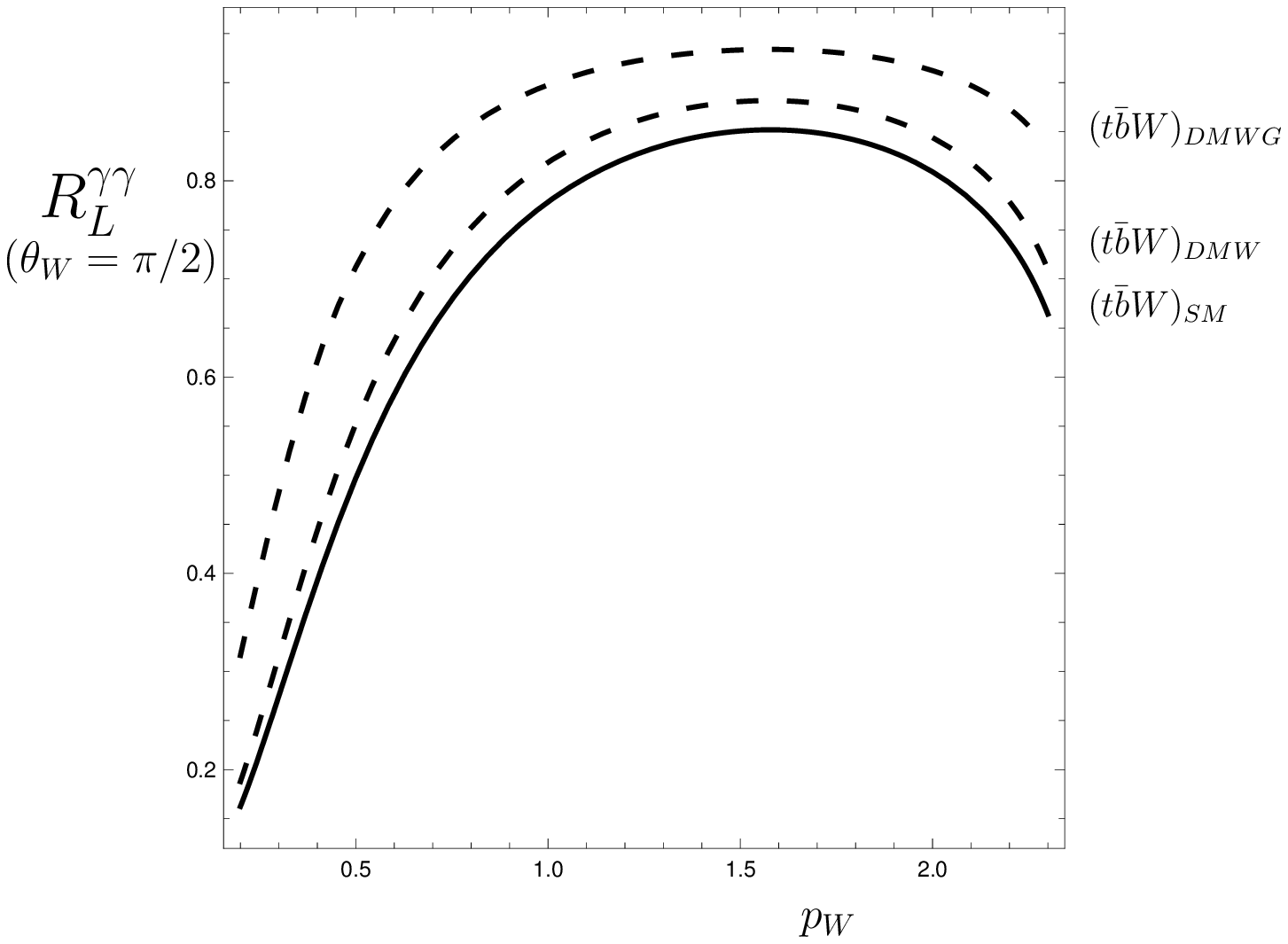 , height=9.cm}
\]\\
\vspace{-1cm}
\caption[1] {$\gamma\gamma\to W_L t\bar b$ ratio for 2 cases of Dark Matter final state interactions.}
\end{figure}
\clearpage


\begin{thebibliography}{99}

%
\bibitem{eettZ} F.M. Renard, arXiv: 1803.10466.
%
\bibitem{ggttZ} F.M. Renard, arXiv: 1805.06379.

%
\bibitem{equiv}
J.M.Cornwall, D.N.Levim and G.Tiktopoulos, Phys. Rev.D10(1974)1145 ;
D11(1975) 972E; C.E.Vayonakis, Lett. Nuovo Cimento17(1976) 383;
B.W.Lee, C.Quigg and H.Thacker, Phys. Rev.D16(1977) 1519 ;
M.S.Chanowitz and M.K.Gaillard, Nucl. Phys.B261(1985) 379;
M.S.Chanowitz, Ann.Rev.Nucl.Part.Sci.38(1988)323;
G.J.Gounaris, R.Koegerler and H.Neufeld, Phys. Rev.D34(1986) 3257.
%
\bibitem{trcomp}  G.J. Gounaris and F.M. Renard,
arXiv: 1611.02426.
%
\bibitem{CSMrev}  F.M. Renard, arXiv: 1708.01111.
%
\bibitem{revDM} B. Penning, arXiv: 1712.01391. We also thank Mike Cavedon
for interesting informations about this subject.  
%
\bibitem{comp}  H. Terazawa, Y. Chikashige and K. Akama, \pr{D15}{480}{1977};
for other references see
H. Terazawa and M. Yasue, Nonlin.Phenom.Complex Syst. {\bf19},1(2016);
\jmp{5}{205}{2014}.
%
\bibitem{Hcomp2} D.B. Kaplan and H. Georgi, \pl{136B}{183}{1984}.
%
\bibitem{Hcomp3} K. Agashe, R. Contino and A. Pomarol, \np{B719}{165}{2005}; hep/ph 0412089.
%
\bibitem{Hcomp4} G. Panico and A. Wulzer, Lect.Notes Phys. {\bf 913},1(2016).
%
\bibitem{partialcomp} R. Contino, T. Kramer, M. Son and R. Sundrum,
J. High Energy Physics {\bf 05}(2007)074.

%
\bibitem{DMmass} F.M. Renard, arXiv: 1712.05352.
%
\bibitem{DMexch} F.M. Renard, arXiv: 1801.10369.
%
\bibitem{ee} G. Moortgat-Pick et al, Eur. Phys. J.C75, 371 (2015), arXiv: 1504.01726. 
%
\bibitem{Contino} R. Contino et al, arXiv: 1606.09408. 
%
\bibitem{Richard}  F. Richard, arXiv: 1703.05046.
%
\bibitem{gammagamma} V.I. Telnov, Nucl.Part.Phys.Proc. {\bf 273}(2016)219.

\end{thebibliography}
\end{document}